\documentclass[aps,twocolumn,superscriptaddress,a4paper,floatfix]{revtex4-2}
\usepackage[latin1]{inputenc}
\usepackage{graphicx}
\usepackage{amsmath,amssymb}
\usepackage{hyperref}

\newcommand \beq{\begin{eqnarray}}
\newcommand \eeq{\end{eqnarray}}

\begin{document}

\title{Asymmetry and nonlinearity of current-bias characteristics in superfluid-normal state junctions of weakly-interacting Bose gases  }

\author{Shun Uchino}
\affiliation{Advanced Science Research Center, Japan Atomic Energy Agency, Tokai 319-1195, Japan}

%\date{\pdfdate}

\begin{abstract}
We uncover current-bias characteristics of superfluid-normal state junctions with weakly-interacting Bose gases.
It is shown that in the presence of a chemical potential bias the characteristics can  strongly be asymmetric for origin.
The salient feature  that is absent in the fermionic counterpart
arises from a tunneling process associated with a condensate and a bosonic Andreev reflection process.
It turns out that such processes are intrinsically nonlinear  and therefore do not obey Ohm's law even at a low bias.
In addition, the remaining processes are found to obey Ohm's law and become dominant for transport driven by a temperature bias.

\end{abstract}

%\pacs{
%}

\maketitle

Recent years have witnessed spectacular development of  atomtronics~\cite{amico2021,amico2021atomtronic}, where
quantum transport phenomena with ultracold atomic gases have been studied under various geometries such as quantum point contact~\cite{krinner2017},
planar junction~\cite{PhysRevLett.126.055301,luick2020},
and junction on spin space~\cite{ono2021}.
Owing to controllability of ultracold atomic gases, one can utilize  quantum states whose
realizations are elusive with typical condensed matter systems including metals, semiconductors, and superconductors.
A weakly-interacting Bose-Einstein condensate (BEC) is the stellar example in that
our understanding on it has significantly deepened since the first realization with ultracold atomic gases~\cite{pethick,pitaevskii2016}.
When it comes to quantum transport of the BECs, Josephson and dissipative currents passing through barriers~\cite{PhysRevLett.95.010402,levy2007,PhysRevLett.106.025302,valtolina2015,PhysRevLett.120.025302}
and tubes~\cite{PhysRevA.93.063619,PhysRevLett.123.260402} have already been measured. 

In quantum transport, it is known that 
junction systems in which different quantum states are attached play a crucial role in condensed matter physics.
A celebrated example is the p-n junction,
which has a direct relevance to electronics and  has been applied for semiconductor devices such as  diodes and transistors~\cite{anderson}.
Another famous example, which is related to the present work, is the superconductor-normal metal (or semiconductor) junction~\cite{tinkham}.
In this system, one can observe the current-bias characteristics that originate from suppression of the quasi-particle transport process~\cite{mahan2013} and 
presence of the effective pair transport process due to the Andreev reflection~\cite{nazarov}.
Especially, the latter process is associated with the conversion between particle and hole, which is imprinted as particle-hole mixing in superconductors described by Bogoliubov-de-Gennes Hamiltonian. 
Since weakly-interacting BECs at low temperature are described by the Bogoliubov Hamiltonian that has
the $2\times2$ matrix structure analogous to one in superconductors, 
the existence of bosonic Andreev reflections in such BECs is expected~\cite{PhysRevLett.102.180405,zapata2011,PhysRevD.98.124043}.
However, transport characteristics of a superfluid-normal state (SN) junction composed of weakly-interacting bosons and
the role of the bosonic Andreev reflection process in terms of mesoscopic transport have yet to be understood.

\begin{figure}[htbp]
\centering
\includegraphics[width=8.7cm]{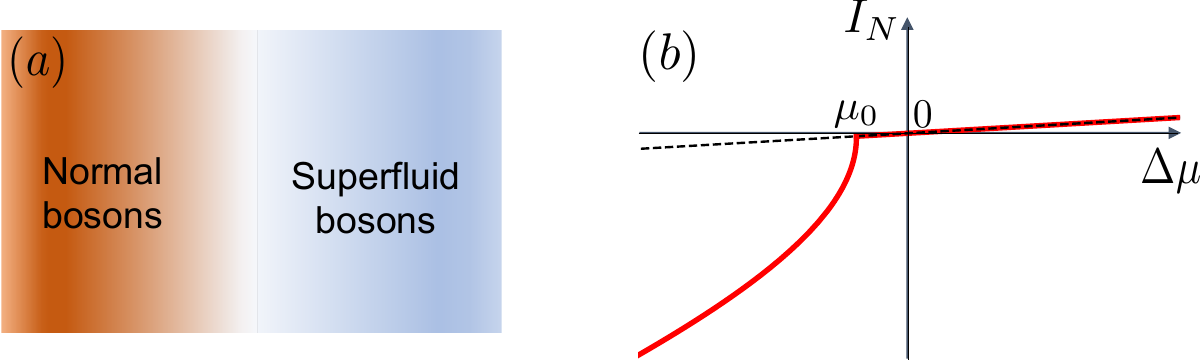}
\caption{\label{fig1} (a) Superfluid-normal state junction of a weakly-interacting Bose gas. We focus on a situation that the transport properties can be described by
the tunneling Hamiltonian formalism.
(b) Typical particle current behavior as a function  of a chemical potential bias $\Delta\mu$ in the corresponding junctions  (solid red curve). 
The characteristic deviates from  symmetric one (dashed line) 
due to presences of tunneling process associated with a condensate and bosonic Andreev reflection process,
both of which emerge for $\Delta\mu<\mu_0$ with chemical potential of noninteracting bosons $\mu_0$.
} 
\end{figure}
In this paper, we microscopically investigate transport characteristics of SN junctions made of  weakly-interacting Bose gases (Fig.~1~(a)).
By analyzing typical situations such as planar junction and point contact via the tunneling Hamiltonian formalism, we demonstrate that unlike the superconductor-normal metal junctions,
 the current-chemical potential bias characteristics show an asymmetry for origin (Fig.~1~(b)),
which is significant in the vicinity of the critical temperature of a BEC.   
This notable  feature of the bosonic junction is ascribed by presences
of a tunnel coupling between condensate and normal boson and bosonic Andreev reflection process, both of which remarkably
do not obey Ohm's law.
Due to chemical potential bias ($\Delta\mu$) and tunnel-coupling dependences, the former is found to be more dominant than the latter over a low-bias regime in which
$\Delta\mu$ is close to the chemical potential of noninteracting bosons.
We also show that the remaining processes that appear  in the corresponding fermionic system are conventional in that an Ohmic response is obtained in the presence of a chemical potential bias and become dominant in transport induced by a temperature bias.

We consider a two-terminal configuration where the left (right) reservoir is filled by
normal (superfluid) bosons and the coupling between reservoirs can be captured with  the following tunneling term:
\beq
H_T=\sum_{\mathbf{p,k}}t_{\mathbf{p,k}}b^{\dagger}_{\mathbf{p},R}b_{\mathbf{k},L}+h.c.
\eeq
Here $t_{\mathbf{p,k}}$ is the tunneling amplitude, and $b_{\mathbf{k},L}(b^{\dagger}_{\mathbf{p},R})$ the annihilation (creation) operator of a boson with momentum $\mathbf{k}$ ($\mathbf{p}$)
in the left (right) reservoir.
In order to obtain comparable results with superconductor-normal metal junctions in which the Bogoliubov-de-Gennes approach is applicable,
we focus on  a situation that  an interaction between bosons in each reservoir is short range and weakly repulsive.
In such a situation, the superfluid reservoir at low temperature and the normal reservoir can be described by the Bogoliubov theory
and the Hartree-Fock (HF) theory~\cite{pitaevskii2016}, respectively.

With these understandings, we now evaluate
the particle current in SN junctions of bosons. By taking the convention that
a particle flow from left to right reservoir is positive, the particle current at time $\tau$ is expressed as
\beq
I_N=2\sum_{\mathbf{p,k}}\text{Re}[t_{\mathbf{p,k}}G^<_{\mathbf{k,p},LR}(\tau,\tau)],
\label{eq:particle}
\eeq
where $G^<_{\mathbf{k,p},LR}(\tau,\tau')=-i\langle b^{\dagger}_{\mathbf{p},R}(\tau')b_{\mathbf{k},L}(\tau)\rangle$
is lesser Green's function~\footnote{We note that  since the particle current does not depend on time, its temporal argument is omitted.}.
We are especially interested in mesoscopic transport induced by chemical potential or temperature bias between reservoirs.
Therefore, we turn to adapt the Keldysh formalism~\cite{rammer2007,kamenev2011}, which provides an efficient program for calculation of mesoscopic currents.
In addition, since in the SN system an AC component is absent in the presence of a static bias unlike superfluid-superfluid junctions~\cite{PhysRevA.64.033610,PhysRevResearch.2.023284},
it is also convenient to work in the frequency space, where the current is expressed as
\beq
I_N=\int_{-\infty}^{\infty} d\omega I_N(\omega),
\eeq 
with the integrand $I_N(\omega)$ with respect to frequency $\omega$.
By solving the Dyson equation for lesser Green's function with the help of the Langreth rule~\cite{rammer2007}, we find that
the current including all processes up to infinite order in tunneling amplitude is obtained as~\cite{suppl}
\beq
I_N=I_{c}+I_1+I_2+I_3+I_A.
\label{eq:mass}
\eeq
The expression above shows that the particle current consists of five different processes.
First, $I_{c}$ is unique to the present system being absent in the superconducting junction case
and describes a process in which a particle in the normal phase is converted into a condensate in the superfluid phase.
We point out that this process resembles a conversion process between condensate and Bogoliubov mode appearring in the superfluid-superfluid junction~\cite{PhysRevA.64.033610,PhysRevA.100.063601,PhysRevResearch.2.023284,PhysRevResearch.2.023340,PhysRevLett.126.055301,PhysRevResearch.3.043058}. 
In such a superfluid junction, the presence of the conversion process between condensate and normal boson (Bogoliubov mode) is inevitable for
tunneling where the momentum conversation is broken.
In addition, $I_1$, $I_2$, and $I_3$ are processes, which are also present in the fermionic case,
describe normal particle transfer between reservoirs, transfer of a particle with creation or annihilation of 
pairs, and a process in which a particle in the normal phase is converted into a hole in the superfluid phase, respectively~\cite{PhysRevB.25.4515,PhysRevB.54.7366}.
Finally, $I_A$ is the bosonic Andreev reflection process where a particle in the normal phase is reflected as a hole.

In order to see a qualitative difference between bosonic and fermionic systems in a simple manner,
we first look at the planar junction case in which the current analysis up to second order in tunneling amplitude
is reasonable. In this case, the particle current passing through the junction  is expressed as
\beq
I_N\approx I_{c}+I_1.
\eeq
Here, $I_2$, $I_3$, $I_A$ are extinguished, since these
are composed of the higher-order processes in tunneling amplitude~\cite{suppl}.
Whereas $I_1$ is the only contribution in the tunneling junction of the fermionic system, 
the contribution of $I_{c}$ also appears in the present bosonic system.
The presence of the additional contribution and the difference of quasi-particle excitations between bosons and fermions give rise to a
nontrivial current response.
To demonstrate this, we focus on the particle current in a low chemical-potential bias  ($\Delta\mu$) regime where the low-energy excitations play dominant roles~\cite{PhysRevA.64.033610,PhysRevResearch.2.023284}.
In this case, we find that  $I_1$ in the bosonic junction obeys  Ohm's law and therefore its differential conductance $G_1$ is independent of $\Delta\mu$. 
This implies that  $I_1$ component in the current-bias curve is symmetric for origin in a similar manner to the corresponding fermionic junction.
We also point out that  $G_1$ shows a peculiar temperature dependence depending on the temperature regime of the normal bosons in which the chemical potential of noninteracting bosons
$\mu_0$ is a key parameter within the HF theory;
$G_1$ is proportional to $T^{7/2}$ near the critical temperature of BECs $T_c$ where $\mu_0\to0^{-}$, whilst 
at high temperature where $\mu_0/T\to-\infty$
 it is suppressed by an exponential factor whose argument is proportional to  $\mu_0/T$~\cite{suppl}.
Especially, the fractional temperature dependence near $T_c$ reflects on the density of states in normal bosons $\rho_N(\omega)\propto\sqrt{\mu_0+\omega}$.
Such a behavior in $I_1$ is  different from the fermionic junction in which the quasi-particle current as a function of $\Delta\mu$ is suppressed by the factor $e^{-\Delta/T}$ 
below the superconducting gap $\Delta$ and shows the abrupt enhancement near $\Delta$~\cite{tinkham}.
The difference between bosonic and fermionic junctions arises from facts that the quasi-particle excitation in the fermions has the gap while
that in the bosons is gapless, and frequency dependence of density of states in normal fermions is ignorable owing to the Fermi level while that in normal bosons is non-negligible.

On the other hand, $I_{c}$ being inherent to the bosonic junction shows an asymmetric behavior in $\Delta\mu$,
since it vanishes for $\Delta\mu>\mu_0.$
This property arises from the fact that  $I_{c}$ is proportional to  $\rho_N(-\Delta\mu)$, which becomes zero unless $\mu_0-\Delta\mu>0$.
Physically, this means that for $\mu_0-\Delta\mu<0$
the conversion between the condensation element and normal boson does not occur,
as a particle in the normal phase cannot find the corresponding condensation element.
Notice that for the low-$\Delta\mu$ regime $I_{c}$ is absent at high temperature where  $\mu_0$ takes a large negative value.
In addition, the appearance of the factor $\rho_N(-\Delta\mu)$ implies that  $I_{c}$ is intrinsically nonlinear and has a square root dependence in $\Delta\mu$.
The presence of the nonlinear curve due to $I_c$ is peculiar to the bosonic SN junction and  is absent 
in the superfluid-superfluid junction of bosons~\cite{PhysRevA.64.033610,PhysRevA.100.063601,PhysRevResearch.2.023284,PhysRevResearch.2.023340,PhysRevLett.126.055301,PhysRevResearch.3.043058}. 

Now that the particle current in the tunneling regime has been demonstrated, we turn to look at
the higher-order tunneling effects. To see this, we consider the single-mode point-contact system in which
tunneling occurs between the single points in each reservoir\cite{PhysRevB.54.7366,PhysRevB.84.155414,husmann2015,PhysRevResearch.2.023284,ono2021,suppl}.
For the point-contact system,
higher-order tunneling processes are generated by controlling the gate potential applied to constriction~\cite{krinner2017} or increasing
an atom-impurity potential via the Feshbach resonance~\cite{ono2021}, and therefore,
all the contributions in the right hand side of Eq.~\eqref{eq:mass} are facilitated. 
The detailed calculation based on the Keldysh formalism reveals that the qualitative change of $I_{c}$ compared to the planar junction case does not occur in that
$I_c$ is proportional to $\rho_N(-\Delta\mu)$, whereas the higher-order tunneling processes give an impact on its proportional coefficient~\cite{suppl}.
Thus, it follows that as in the planar junction case  $I_{c}$ in the point contact case is relevant near $T_c$.
In addition, we find $I_1\approx G_1\Delta\mu$ for low $\Delta\mu$, where $G_1$ is independent of  $\Delta\mu$ similar to the planar junction.
Due to the difference in $t_{\mathbf{p,k}}$ between the planar junction and the point contact, however,
$G_1$ in the point contact  is found to be proportional to $T^{3/2}$ near $T_c$, which is different from one in the planar junction~\cite{suppl}.
With increasing the coupling between reservoirs, $I_2$ and $I_3$ whose leading order contributions are respectively proportional to fourth order and sixth order in $t_{\mathbf{p,k}}$
are generated. By performing an explicit calculation for low $\Delta\mu$, we find that  $I_2$ and $I_3$ also obey Ohm's law where the corresponding
conductances $G_2$ and $G_3$ are proportional to $T^{3/2}$ near $T_c$ and are exponentially suppressed at high temperature as in $G_1$~\cite{suppl}.

In contrast, $I_A$ whose leading contribution is proportional to the fourth order in $t_{\mathbf{p,k}}$ turns out to possess a non-Ohmic character.
To see this, we point out that the integrand in $I_A$  contains the product of densities of states in particle and hole branches of normal bosons, that is,
\beq
\rho_N(\omega-\Delta\mu)\rho_N(-\omega-\Delta\mu).
\label{eq:dos}
\eeq
Thus, it follows that in order for $I_A$ to be nonzero, there must be  frequencies that simultaneously satisfy  $\mu_0+\omega-\Delta\mu>0$ and $\mu_0-\omega-\Delta\mu>0$.
This indicates that $I_A$ can be nonzero for $\Delta\mu<\mu_0$ and contribute to low-energy transport near $T_c$ similar to $I_{c}$.
In addition, the fact that a particle is reflected as a hole in the Andreev reflection process implies that the integrand in $I_A$ also contains the factor
\beq
n_L(\omega-\Delta\mu)-n_L(\omega+\Delta\mu),
\label{eq:andreev}
\eeq
where $n_L$ is the Bose distribution function in (normal) reservoir $L$.
 Equation~\eqref{eq:andreev} is linear in $\Delta\mu$ for the low bias regime. 
However, Ohm's law is prohibited in $I_A$ due to the dependence in densities of states (Eq.~\eqref{eq:dos}), which introduces an additional $\Delta\mu$ dependence in the integrand of $I_A$.
An explicit numerical analysis  shows that  
$I_A\propto (\Delta\mu-\mu_0)^2\theta(\mu_0-\Delta\mu)$ for small $\Delta\mu$ and $\mu_0$~\cite{suppl}. 
This is in sharp contrast to the Andreev reflection contribution in the superconductor where 
the Andreev current obeys Ohm's law for the low-bias regime~\cite{PhysRevB.25.4515,PhysRevB.54.7366}. 
We also point out that up to this order of the expression, $I_A$ has no temperature dependence as in $I_c$.
Since the leading order contribution in $I_c$ is the second order in $t_{\mathbf{p,k}}$ and $I_c\propto\sqrt{\mu_0-\Delta\mu}$, however,
$I_c$ is more dominant than $I_A$ over the low-bias regime.

Besides transport induced by a chemical potential bias, 
 we can discuss one induced sorely by a temperature bias  between reservoirs $\Delta T$.
 Regarding the particle current, it turns out that both $I_{c}$ and $I_A$ become zero.
 The former contribution is absent since the density of states in the normal bosons vanishes at zero frequency.
On the other hand, the latter one is absent due to the fact that the Andreev reflection process
 contains the factor Eq.~\eqref{eq:andreev} in the integrand, which trivially vanishes at $\Delta\mu=0$. 
 In other words, the remaining processes $I_1$, $I_2$, and $I_3$ are generated by $\Delta T$.
 It is then straightforward to show that in the presence of small $\Delta T$ these contributions are proportional to $\Delta T$.
 Furthermore, we discuss the heat current $I_Q$ induced by $\Delta T$.
By using lesser Green's function, it is expressed as
\beq
I_Q=2\sum_{\mathbf{p,k}}\lim_{\tau'\to\tau}\text{Re}\Big[it_{\mathbf{p,k}}\partial_{\tau}G^<_{\mathbf{k,p},LR}(\tau,\tau') \Big].
\label{eq:heat}
\eeq
As in the case of the particle current, the heat current can efficiently be calculated in the frequency space.
Then, $i\partial_{\tau}$ in Eq.~\eqref{eq:heat} is replaced by $\omega$ in the frequency representation and 
the integrand of Eq.~\eqref{eq:heat} is given by one of Eq.~\eqref{eq:particle} multiplied by $\omega$, i.e., $I_Q=\int_{-\infty}^{\infty} d\omega\omega I_N(\omega)$.
The direct evaluation of $I_Q$ shows that the contributions coming from the conversion process between condensate and normal boson and the Andreev reflection process
vanish, and the remaining processes corresponding to $I_1$, $I_2$, and $I_3$ in the particle current become dominant.
In addition, we can show that for the low-$\Delta T$ regime, $I_Q$ is proportional to $\Delta T$, implying the presence of Fourier's law.

We now discuss possible realizations of the bosonic SN junctions with ultracold atomic gases.
Unlike condensed-matter setups where superconductor and normal metal consist of different materials,
 junction systems in cold-atom experiments are made of an atomic gas originally trapped in one~\cite{krinner2017}.
By considering that the critical temperature of the BEC depends on an atomic density,
in order to verify the predictions discussed in this work,
one has to make a two-terminal system where the chemical potential difference between reservoirs is small
and density difference between reservoirs is sufficiently large such that the thermodynamic properties of the left and right reservoirs
are described by the HF and the Bogoliubov theories, respectively.
To achieve this, one can use a repulsive external potential, which is only added in the left reservoir and 
compensates the chemical potential difference arising from the large density difference between reservoirs~\cite{price}.
Another way to make the similar system is to use the synthetic-reservoir technique available in ultracold atomic gases with internal degrees of freedom.
There, the hyperfine or nuclear spin index of each atom is regarded as the reservoir index, and
 tunneling and chemical-potential shift between reservoirs can be introduced as coupling and energy-level shift between spins, respectively.
Such a technique has already been implemented to confirm the Josephson effect~\cite{PhysRevLett.105.204101} and point contact transport~\cite{ono2021}.

For realizations of the SN junctions, it is instructive to point out that due to the critical fluctuation the HF theory is broken down in the very vicinity of $T_c$,
where $\mu_0$ approaches to zero. By using the Ginzburg criterion, such a fluctuation is important around the critical chemical potential $\mu_c=(4\pi a)^2MT_c^2$
with the mass of an atom $M$ and the $s$-wave scattering length $a$~\cite{pitaevskii2016}.
In order to see the existence of the regime where the theoretical analysis addressed in this work is justified in a concrete way,
we consider a system of $^{87}$Rb atoms and assume that the atomic densities in the normal and superfluid reservoirs are respectively of the order of $10^{11}$cm$^{-3}$ and $10^{13}$cm$^{-3}$.
In this case, $T_c\sim 10^{-9}$K and $\mu_c\sim 10^{-11}$K in the normal reservoir, while $T_c\sim10^{-8}$K in the superfluid reservoir.
Thus, if the temperature of the system is $T\sim10^{-9}$K,
we can reach  the regime in which the superfluid reservoir is described by the Bogoliubov theory and
normal reservoir is done by the HF theory while avoiding the critical fluctuation.

To summarize, we addressed the current-bias characteristics of the bosonic SN systems.
Owing to the presence of the tunneling process inherent to the bosonic systems and the excitation properties,
 the transport characteristics turn out to be drastically different from the  fermionic counterparts.
Especially, we revealed that in the vicinity of the critical temperature of a BEC, the current-chemical potential bias characteristics can
be nonlinear due to the conversion process between condensate and normal boson and the bosonic Andreev process.
At the same time,  such processes yielding the nonlinear character are found to be irrelevant for transport induced by a temperature bias.
 
The results shown in this paper may encourage a variety of related works.
An analysis in terms of the scattering formalism with the Bogoliubov equation is an interesting direction for future research.
Such an analysis may play a key role in connecting the tunneling process such as the conversion process discussed in this work with the so-called quantum condensation 
and evaporation~\cite{PhysRevLett.16.1191,anderson1969,PhysRevLett.52.1528,PhysRevLett.75.2510,PhysRevA.94.023622}.
In addition, the formalism explored in this work can be applied to various BEC systems discussed in ultracold atomic gases.
Weakly-interacting Bose gases with spin could be one of the  systems. There, the current-bias characteristics may further be nontrivial, 
since such BECs possess a variety of Bogoliubov excitations depending on the phases realized~\cite{RevModPhys.85.1191}.
Similarly, junction systems of the binary mixtures with or without spin-orbit coupling are promising in terms of
comparable superconducting systems~\cite{zhai2021}.
Furthermore, the SN junction in two-component Fermi gases may be interesting in terms of Bardeen-Cooper-Schrieffer (BCS)-BEC crossover~\cite{zwerger}. 
There, questions would be how evolution of the transport characteristics as a function of the $s$-wave scattering length
and crossover between fermionic and bosonic Andreev reflections are.
Another promising direction may be to look at optical lattice systems in which
an effect of commensurability is significant near the Mott-insulator transition~\cite{sachdev2011}.

\section*{acknowledgment}
The author is supported by MEXT Leading Initiative for Excellent Young Researchers,
JSPS KAKENHI Grant No.~JP21K03436, and Matsuo Foundation.

\bibliographystyle{apsrev4-1}
%\bibliography{reference}
%

\pagebreak
\widetext
\begin{center}
\textbf{\large Supplemental Materials}
\end{center}
%%%%%%%%%% Merge with supplemental materials %%%%%%%%%%
%%%%%%%%%% Prefix a "S" to all equations, figures, tables and reset the counter %%%%%%%%%%
\setcounter{equation}{0}
\setcounter{figure}{0}
\setcounter{table}{0}
\setcounter{page}{1}
\makeatletter
\renewcommand{\theequation}{S\arabic{equation}}
\renewcommand{\thefigure}{S\arabic{figure}}
\renewcommand{\bibnumfmt}[1]{[S#1]}
\renewcommand{\citenumfont}[1]{S#1}

\section{Weakly-interacting Bose gases}
We consider a  system of a spinless Bose gas with mass $M$ that undergoes an $s$-wave scattering.
The grand canonical Hamiltonian of the system is given by
\beq
H=\int d^3x \phi^{\dagger}(\mathbf{x})\Big[-\frac{\nabla^2}{2M} -\mu\Big]\phi(\mathbf{x})+\frac{g}{2}\int d^3x\phi^{\dagger}(\mathbf{x})\phi^{\dagger}(\mathbf{x})\phi(\mathbf{x})\phi(\mathbf{x}),
\label{eq:hamiltonian}
\eeq
where $\phi(\mathbf{x})$ $(\phi^{\dagger} (\mathbf{x})$) is the annihilation (creation) operator of a boson at position $\mathbf{x}$,
$g$  the coupling constant of the two-body interaction, $\mu$ the chemical potential, and 
we note that we adopt units of $\hbar=k_B=1$.
Since the interaction between bosons is assumed to be weak, the coupling constant $g$ is related to the $s$-wave scattering length $a$ as follows~\cite{pitaevskii2016}:
\beq
g=\frac{4\pi a}{M}.
\eeq

\subsection{Bogoliubov theory}
We consider the low-temperature limit of the Bose gas, where almost all of the bosons condense at $\mathbf{k}=\mathbf{0}$.
In such a limit, $\mathbf{k}=\mathbf{0}$ component in $\phi(\mathbf{x})$ gives a dominant contribution.
In the Bogoliubov theory, the dominant component is replaced by $c$-number and 
an effect of the fluctuation part ($\mathbf{k}\ne\mathbf{0}$) is incorporated up to second order in the Hamiltonian~\cite{pitaevskii2016}.
By using this prescription, Eq.~\eqref{eq:hamiltonian} is approximated as
\beq
H_{\text{B}}=\sum_{\mathbf{k}}(\epsilon_{\mathbf{k}}+\mu)b^{\dagger}_{\mathbf{k}}b_{\mathbf{k}}
+\frac{\mu}{2}\sum_{\mathbf{k}}[b^{\dagger}_{\mathbf{k}}b^{\dagger}_{\mathbf{-k}} +b_{\mathbf{k}}b_{\mathbf{-k}}],
\label{eq:bogoliubov}
\eeq
where $\epsilon_{\mathbf{k}}=k^2/(2M)$ and $b_{\mathbf{k}}$ is the annihilation operator of a boson with momentum $\mathbf{k}$.
We note that in Eq.~\eqref{eq:bogoliubov} we drop the constant term arising from the $c$-number field.
Actually, such a term determines the following relation between the $c$-number field $v$ and the chemical potential:
\beq
\mu=vg.
\eeq
Equation~\eqref{eq:bogoliubov} can be diagonalized by the Bogoliubov transformation:
\beq
b_{\mathbf{k}}=u_k\beta_{\mathbf{k}}-v_k\beta^{\dagger}_{-\mathbf{k}},
\label{eq:nambu}
\eeq
with
\beq
&&u_k=\sqrt{\frac{\epsilon_{\mathbf{k}}+\mu+E_{\mathbf{k}} }{2E_\mathbf{k}}},\\
&&v_k=\sqrt{\frac{\epsilon_{\mathbf{k}}+\mu-E_{\mathbf{k}} }{2E_\mathbf{k}}},\\
&&E_{\mathbf{k}}=\sqrt{\epsilon_{\mathbf{k}}(\epsilon_{\mathbf{k}}+2\mu) }.
\eeq
Namely, in the representation of $\beta_{\mathbf{k}}$, Eq.~\eqref{eq:bogoliubov} is expressed as
\beq
H_B=\sum_{\mathbf{k}}E_{\mathbf{k}}\beta^{\dagger}_{\mathbf{k}}\beta_{\mathbf{k}}.
\eeq

We now introduce a couple of Green's functions used in the Keldysh formalism.
Especially, in the presence of the BEC, it is convenient to introduce the Nambu spinor:
\beq
\hat{b}_{\mathbf{k}}=\begin{pmatrix}
b_{\mathbf{k}} \\
b^{\dagger}_{-\mathbf{k}}
\end{pmatrix}.
\eeq
By using the Nambu spinor, we can introduce $2\times2$ matrix contour-ordered Green's function as follows:
\beq
\hat{G}_{\mathbf{p} }(\tau_{a},\tau'_b)=-i\langle T_c[\hat{b}_{\mathbf{k}}(\tau)\hat{b}^{\dagger}_{\mathbf{k}}(\tau') ] \rangle,
\eeq
where $T_c$ is the contour ordering.
We now apply the formalism above to the Bogoliubov theory, which can be done by the replacement $H\to H_\text{B}$ in Eq.~\eqref{eq:nambu}.
In the frequency space, retarded Green's functions are obtained as
\beq
\hat{g}^R_{\mathbf{k}}(\omega)&&=-i\int_{-\infty}^{\infty}d\tau\theta(\tau)\langle \hat{b}_{\mathbf{k}}(\tau),\hat{b}^{\dagger}_{\mathbf{k}}(0) ]\rangle\nonumber\\
&&=\begin{pmatrix}
\frac{u_k^2}{\omega-E_{\mathbf{k}}+i0^+ }-\frac{v_k^2}{\omega+E_{\mathbf{k}}+i0^+ } & \frac{u_kv_k}{\omega-E_{\mathbf{k}}+i0^+ }-\frac{u_kv_k}{\omega+E_{\mathbf{k}}+i0^+ } \\
\frac{u_kv_k}{\omega-E_{\mathbf{k}}+i0^+ }-\frac{u_kv_k}{\omega+E_{\mathbf{k}}+i0^+ }& \frac{v_k^2}{\omega-E_{\mathbf{k}}+i0^+ }-\frac{u_k^2}{\omega+E_{\mathbf{k}}+i0^+ } 
\end{pmatrix}.
\eeq
Advanced Green's function $\hat{g}^A$ can be determined with $\hat{g}^A_{\mathbf{k}}(\omega)=[g^R_{\mathbf{k}}(\omega)]^{\dagger} $.
In addition,  lesser Green's function is obtained as
\beq
\hat{g}^<_{\mathbf{k}}(\omega)&&=-i\int_{-\infty}^{\infty}d\tau e^{i\omega\tau}
\begin{pmatrix}
\langle  b^{\dagger}_{\mathbf{k}}(0)b_{\mathbf{k}}(\tau) \rangle & \langle  b_{-\mathbf{k}}(0)b_{\mathbf{k}}(\tau) \rangle \\
\langle  b^{\dagger}_{\mathbf{k}}(0)b^{\dagger}_{-\mathbf{k}}(\tau) \rangle & \langle  b_{\mathbf{k}}(0)b^{\dagger}_{\mathbf{k}}(\tau) \rangle 
\end{pmatrix}\nonumber\\
&&=\begin{cases}
-\frac{2\pi\mu i}{g}\delta(\omega)\begin{pmatrix}1&1\\1&1 \end{pmatrix} \ \ \ (\mathbf{k}=\mathbf{0}) \\
\begin{pmatrix}
-2\pi in(\omega)[u_k^2\delta(\omega-E_{\mathbf{k}}) -v_k^2\delta(\omega+E_{\mathbf{k}})] & 2\pi in(\omega)u_kv_k[\delta(\omega-E_{\mathbf{k}}) -\delta(\omega+E_{\mathbf{k}})] \\
 2\pi in(\omega)u_kv_k[\delta(\omega-E_{\mathbf{k}}) -\delta(\omega+E_{\mathbf{k}})] & 2\pi in(\omega)[u_k^2\delta(\omega+E_{\mathbf{k}}) -v_k^2\delta(\omega-E_{\mathbf{k}})] 
\end{pmatrix} \ \ \ (\mathbf{k}\ne\mathbf{0})
\end{cases}
\eeq
where $n(\omega)=\frac{1}{e^{\omega/T}-1 }$ is the Bose distribution function.

For the sake of the latter discussion, we also express local Green's functions $\hat{g}=\sum_{\mathbf{k}}\hat{g}_{\mathbf{k}}$.
The retarded component is calculated as~\cite{PhysRevResearch.3.043058}
\beq
\hat{g}^R(\omega)=\begin{pmatrix}
\Phi(\omega)-i\pi\rho(\omega) & \Psi(\omega)-i\pi\sigma(\omega)  \\
 \Psi(\omega)-i\pi\sigma(\omega)   & \Phi(-\omega)+i\pi\rho(-\omega) 
\end{pmatrix},
\eeq
where 
\beq
\Phi(\omega)&&=-\frac{\sqrt{2}\mu^2 }{c^3\pi^2}
\Big[\frac{c\Lambda}{\sqrt{2}\mu }+\frac{\pi(\bar{\omega}-1 )}{4\sqrt{1+\bar{\omega}^2}}\Big\{\sqrt{1+\sqrt{1+\bar{\omega}^2}}-
\sqrt{1-\sqrt{1+\bar{\omega}^2}}\theta(1-\sqrt{1+\bar{\omega}^2}) \Big\}\nonumber\\
&&-\frac{\pi\bar{\omega}^2}{4\sqrt{1+\bar{\omega}^2}}\Big\{\frac{1}{\sqrt{1+\sqrt{1+\bar{\omega}^2}}}-
\frac{\theta(1-\sqrt{1+\bar{\omega}^2})}{\sqrt{1-\sqrt{1+\bar{\omega}^2}}}
\Big\}
\Big],
\eeq
\beq
\Psi(\omega)=\frac{\sqrt{2}\mu^2}{c^3\pi^2}\Big[ \frac{\pi}{4\sqrt{1+\bar{\omega}^2}}\Big\{\sqrt{1+\sqrt{1+\bar{\omega}^2}}-
\sqrt{1-\sqrt{1+\bar{\omega}^2}}\theta(1-\sqrt{1+\bar{\omega}^2}) \Big\} \Big],
\eeq
and for $\omega>0$
\beq
\rho(\omega)=\frac{k_0(\epsilon_{k_0}+\mu+E_{k_0})}{4\pi^2c^2(1+\epsilon_{k_0}/\mu)}\theta(\Lambda-k_0),\\
\rho(-\omega)=-\frac{k_0(\epsilon_{k_0}+\mu-E_{k_0})}{4\pi^2c^2(1+\epsilon_{k_0}/\mu)}\theta(\Lambda-k_0),\\
\sigma(\omega)=-\frac{\mu k_0}{4\pi^2c^2(1+\epsilon_{k_0}/\mu)}\theta(\Lambda-k_0),\\
\sigma(-\omega)=\frac{\mu k_0}{4\pi^2c^2(1+\epsilon_{k_0}/\mu)}\theta(\Lambda-k_0).
\eeq
In the above expression, we introduce the cutoff momentum $\Lambda$ and 
\beq
k_0=\sqrt{2M\sqrt{\mu^2+\omega^2}-2M\mu}.
\eeq

In a similar manner, the lesser and greater components are obtained as
\beq
\hat{g}^<(\omega)=-\frac{2\pi\mu i}{g}\begin{pmatrix}
1 & 1\\
1 & 1
\end{pmatrix}
-2\pi in(\omega)\hat{\rho}(\omega),
\eeq
\beq
\hat{g}^>(\omega)=-\frac{2\pi\mu i}{g}\begin{pmatrix}
1 & 1\\
1 & 1
\end{pmatrix}
-2\pi i\{1+n(\omega)\hat{\rho}(\omega),
\eeq
where
\beq
\hat{\rho}(\omega)=\begin{pmatrix}
\rho(\omega) & \sigma(\omega)\\
\sigma(\omega) & \rho(\omega)
\end{pmatrix}.
\eeq

\subsection{Hartree-Fock theory}
We next consider a system with a temperature higher than the critical temperature of the BEC $T_c$.
Such a regime of the weakly-interacting Bose gas can be described by the Hartree-Fock (HF) theory~\cite{pitaevskii2016}.
In this theory, the interaction term is replaced by $\frac{g}{2}\int d^3x\phi^{\dagger}(\mathbf{x})\phi^{\dagger}(\mathbf{x})\phi(\mathbf{x})\phi(\mathbf{x})
\to 2gn\int d^3x\phi^{\dagger}(\mathbf{x})\phi(x)$, where $n$ is the density of the gas.
By using this prescription, Eq.~\eqref{eq:hamiltonian} is approximated as
\beq
H_{\text{HF}}=\sum_{\mathbf{k}}(\epsilon_{\mathbf{k}}+2gn -\mu)b^{\dagger}_{\mathbf{k}}b_{\mathbf{k}}.
\eeq
The chemical potential with the HF theory is obtained as
\beq
\mu=2gn+\mu_0,
\eeq
where $\mu_0$ is the chemical potential of free bosons and is determined from
\beq
n\left(\frac{2\pi}{MT} \right)^{3/2}=\frac{2}{\sqrt{\pi}}\int_0^{\infty}dx\frac{\sqrt{x}}{e^{x-\mu_0/T}-1}.
\eeq
In addition, $T_c$ with the HF theory is same as that of free bosons given by
\beq
T_c=\frac{3.31n^{2/3}}{M}.
\eeq

As in the case of the Bogoliubov theory, Green's functions within the HF theory by the replacement $H\to H_{\text{HF}}$ in Eq.~\eqref{eq:nambu}.
Retarded and advanced Green's functions are obtained as
\beq
\hat{g}^{R/A}_{\mathbf{k}}(\omega)=\begin{pmatrix}
\frac{1}{\omega-\omega_k\pm i0^+} &0\\
0& -\frac{1}{\omega+\omega_k\pm i0^+}
\end{pmatrix},
\eeq
where $\omega_k=\frac{k^2}{2M}+2ng-\mu=\frac{k^2}{2M}-\mu_0$.
In addition, lesser Green's function is obtained as
\beq
\hat{g}^<_{\mathbf{k}}(\omega)=\begin{pmatrix}
 -2\pi i n(\omega)\delta(\omega-\omega_k)&0\\
 0 & 2\pi i n(\omega)\delta(\omega+\omega_k)
\end{pmatrix}.
\eeq
As in the case of the Bogoliubov theory, we now obtain local Green's functions.
Local retarded Green's function is calculated as
\beq
\hat{g}^R(\omega)=\begin{pmatrix}
R(\mu_0+\omega)-i\pi\rho_N(\omega) & 0\\
0 & R(\mu_0-\omega) +i\pi\rho_N(-\omega)
\end{pmatrix},
\eeq
where
\beq
R(\mu_0\pm\omega)=\begin{cases}
-\frac{M}{\pi^2}\Big[\Lambda+\sqrt{2M(\mu_0\pm\omega)}\log\Big|\frac{1-\sqrt{2M(\mu_0\pm\omega)}/\Lambda }{1+\sqrt{2M(\mu_0\pm\omega)}/\Lambda }\Big| \Big], \ \ \ (\mu_0\pm\omega>0) \\
-\frac{M}{\pi^2}\Big[\Lambda-\sqrt{2M|\mu_0\pm\omega|}\arctan\frac{\Lambda}{\sqrt{2M|\mu_0\pm\omega|}} \Big], \ \ \ (\mu_0\pm\omega<0)
\end{cases},
\eeq
\beq
\rho_N(\omega)=\frac{M\sqrt{2M(\mu_0+\omega)}}{2\pi^2}\theta(\mu_0+\omega).
\eeq
The local lesser and greater components are calculated as
\beq
\hat{g}^<(\omega)=-2\pi in(\omega)\hat{\rho}(\omega),
\eeq
\beq
\hat{g}^>(\omega)=-2\pi i\{1+n(\omega)\}\hat{\rho}(\omega), 
\eeq
where
\beq
\hat{\rho}(\omega)= \begin{pmatrix}
\rho_N(\omega) & 0\\
0 & -\rho_N(-\omega)
\end{pmatrix}.
\eeq

\section{Current expressions}
In the tunneling Hamiltonian formalism, the particle current between reservoirs is defined as
\beq
I=-\dot{N}_L(\tau).
\eeq
By using the Heisenberg equation of $N_L$, the current is expressed as
\beq
I&&=-i\sum_{\mathbf{p,k}}t_{\mathbf{p,k}}\langle b^{\dagger}_{\mathbf{p},R}(\tau)b_{\mathbf{k},L}(\tau) \rangle+h.c.\nonumber\\
&&=2\sum_{\mathbf{p.k}} \int_{-\infty}^{\infty}\frac{d\omega}{2\pi}\text{Re}[\hat{t}_{\mathbf{p,k}}\hat{G}^<_{\mathbf{k,p},LR}(\omega)]_{11},
\eeq
where $\hat{G}$ represents full Green's function including the tunneling term in the Hamiltonian, and
we introduce
\beq
\hat{t}_{\mathbf{p,k}}=\begin{pmatrix}
t_{\mathbf{p,k}} & 0\\
0 & \bar{t}_{\mathbf{p,k}}
\end{pmatrix}.
\eeq
Moreover, by using
\beq
\hat{G}^R-\hat{G}^A=\hat{G}^>-\hat{G}^<,
\eeq
and the Langreth rule, i.e., $(AB)^{<(>)}=A^RB^{<(>)}+A^{<(>)}B^A$, 
the current is rewritten as
\beq
I=\int_{-\infty}^{\infty}\frac{d\omega}{2\pi}\Big[t \hat{g}^>_{L,11}(\omega-\Delta\mu)\bar{t} \hat{G}^<_{RR,11}(\omega)-t \hat{g}^<_{L,11}(\omega-\Delta\mu)\bar{t}\hat{G}^>_{RR,11}(\omega)\Big],
\eeq
with the chemical potential difference between reservoirs $\Delta\mu$.
Here, we introduce a shorthand notation that omits the momentum indices of the tunneling matrix and Green's functions and summation over the  momentum variable.
In order to obtain a convenient expression, we look at the Dyson equation for $\hat{G}^{<(>)}$,
\beq
\hat{G}^{<(>)}_{RR}=(\hat{1}+\hat{G}^R_{RL}\hat{t}^{\dagger})\hat{g}^{<(>)}_R(\hat{1}+\hat{t}\hat{G}^A_{LR})+\hat{G}^R_{RR}\hat{t}\hat{g}^{<(>)}_L\hat{t}^{\dagger}\hat{G}^A_{RR}.
\eeq
By using this, we have
\beq
I&&=\int_{-\infty}^{\infty}\frac{d\omega}{2\pi}\Big[t \hat{g}^>_{L,11}(\omega-\Delta\mu)\bar{t}
 (1+\hat{G}^R_{RL,11}(\omega)\bar{t})\hat{g}^<_{R,11}(\omega)(1+t\hat{G}^A_{LR,11}(\omega))\nonumber\\
&&-t \hat{g}^<_{L,11}(\omega-\Delta\mu)\bar{t} 
 (1+\hat{G}^R_{RL,11}(\omega)\bar{t})\hat{g}^>_{R,11}(\omega)(1+t\hat{G}^A_{LR,11}(\omega))\nonumber\\
&&+ t \hat{g}^>_{L,11}(\omega-\Delta\mu)\bar{t}
\{\hat{G}^R_{RL,12}(\omega)t\hat{g}^<_{R,21}(\omega)(1+t\hat{G}^A_{LR,11}(\omega))
+(1+\hat{G}^R_{RL,11}(\omega)\bar{t})\hat{g}^<_{R,12}(\omega)\bar{t}\hat{G}^A_{LR,21}(\omega) \}\nonumber\\
&&- t \hat{g}^<_{L,11}(\omega-\Delta\mu)\bar{t}\{\hat{G}^R_{RL,12}(\omega)t\hat{g}^>_{R,21}(\omega)(1+t\hat{G}^A_{LR,11}(\omega))
+(1+\hat{G}^R_{RL,11}(\omega)\bar{t})\hat{g}^>_{R,12}(\omega)\bar{t}\hat{G}^A_{LR,21}(\omega) \}\nonumber\\
&&+t \hat{g}^>_{L,11}(\omega-\Delta\mu)\bar{t}\hat{G}^R_{RL,12}(\omega)t\hat{g}^<_{R,22}(\omega)\bar{t}\hat{G}^A_{LR,21}(\omega)
-t \hat{g}^<_{L,11}(\omega-\Delta\mu)\bar{t}\hat{G}^R_{RL,12}(\omega)t\hat{g}^>_{R,22}(\omega)\bar{t}\hat{G}^A_{LR,21}(\omega) \nonumber\\
&&+t \hat{g}^>_{L,11}(\omega-\Delta\mu)\bar{t}\hat{G}^R_{RR,12}(\omega)\bar{t}\hat{g}^<_{L,22}(\omega+\Delta\mu)t\hat{G}^A_{RR,21}(\omega)
-t \hat{g}^<_{L,11}(\omega-\Delta\mu)\bar{t} \hat{G}^R_{RR,12}(\omega)\bar{t}\hat{g}^>_{L,22}(\omega+\Delta\mu)t\hat{G}^A_{RR,21}(\omega)
\Big].\nonumber\\
\label{eq:full-current}
\eeq
The expression above contains all the transport processes in the junction.
The contribution containing the condensation element $I_{c}$ can be obtained by substituting the $\mathbf{k}=\mathbf{0}$ element of $\hat{g}^<_R(\omega)$ 
into \eqref{eq:full-current}.
The remaining contributions are obtained by using the $\mathbf{k}\ne\mathbf{0}$ elements of $\hat{g}^<_R(\omega)$ and are given by
\beq
I_1&&=\int_{-\infty}^{\infty}\frac{d\omega}{2\pi}\Big[ t \hat{g}^>_{L,11}(\omega-\Delta\mu)\bar{t}
 (1+\hat{G}^R_{RL,11}(\omega)\bar{t})\hat{g}^<_{nc,R,11}(\omega)(1+t\hat{G}^A_{LR,11}(\omega))\nonumber\\
&&-t \hat{g}^<_{L,11}(\omega-\Delta\mu)\bar{t} 
 (1+\hat{G}^R_{RL,11}(\omega)\bar{t})\hat{g}^>_{nc,R,11}(\omega)(1+t\hat{G}^A_{LR,11}(\omega))\Big],
\eeq
\beq
I_2&&=\int_{-\infty}^{\infty}\frac{d\omega}{2\pi}\Big[ t \hat{g}^>_{L,11}(\omega-\Delta\mu)\bar{t}
\{\hat{G}^R_{RL,12}(\omega)t\hat{g}^<_{nc,R,21}(\omega)(1+t\hat{G}^A_{LR,11}(\omega))
+(1+\hat{G}^R_{RL,11}(\omega)\bar{t})\hat{g}^<_{nc,R,12}(\omega)\bar{t}\hat{G}^A_{LR,21}(\omega) \}\nonumber\\
&&- t \hat{g}^<_{L,11}(\omega-\Delta\mu)\bar{t}\{\hat{G}^R_{RL,12}(\omega)t\hat{g}^>_{nc,R,21}(\omega)(1+t\hat{G}^A_{LR,11}(\omega))
+(1+\hat{G}^R_{RL,11}(\omega)\bar{t})\hat{g}^>_{nc,R,12}(\omega)\bar{t}\hat{G}^A_{LR,21}(\omega) \}\Big],
\eeq
\beq
I_3&&=\int_{-\infty}^{\infty}\frac{d\omega}{2\pi}\Big[
t \hat{g}^>_{L,11}(\omega-\Delta\mu)\bar{t}\hat{G}^R_{RL,12}(\omega)t\hat{g}^<_{nc,R,22}(\omega)\bar{t}\hat{G}^A_{LR,21}(\omega)
-t \hat{g}^<_{L,11}(\omega-\Delta\mu)\bar{t}\hat{G}^R_{RL,12}(\omega)t\hat{g}^>_{nc,R,22}(\omega)\bar{t}\hat{G}^A_{LR,21}(\omega) 
\Big],\nonumber\\
\eeq
\beq
I_A&&=\int_{-\infty}^{\infty}\frac{d\omega}{2\pi}\Big[
t \hat{g}^>_{L,11}(\omega-\Delta\mu)\bar{t}\hat{G}^R_{RR,12}(\omega)\bar{t}\hat{g}^<_{L,22}(\omega+\Delta\mu)t\hat{G}^A_{RR,21}(\omega)\nonumber\\
&&-t \hat{g}^<_{L,11}(\omega-\Delta\mu)\bar{t} \hat{G}^R_{RR,12}(\omega)\bar{t}\hat{g}^>_{L,22}(\omega+\Delta\mu)t\hat{G}^A_{RR,21}(\omega)
\Big],
\eeq
where $\hat{g}_{nc,R}(\omega)$ represents the $\mathbf{k}\ne\mathbf{0}$ contribution in lesser Green's function.
Although the current formulas above contain all order contributions in the tunneling amplitudes,
it is useful to understand the leading order contribution in each process.
To this end, we note that $\hat{G}_{RL(LR)}$ contains at least first order in $t$, while 
$\hat{G}_{LL(RR)}$ starts from zeroth order in $t$.
Thus, it turns out that the leading order contributions of $I_{c}$ and $I_1$ are $t^2$, those of $I_2$ and $I_A$ are $t^4$,
and that of $I_3$ is $t^6$.

\subsection{Planar junction case}
We now consider the planar junction where the current expression up to $t^2$ is concerned. 
In this case, the particle current is obtained as
\beq
I=\sum_{\mathbf{p,k}}\int_{-\infty}^{\infty}\frac{d\omega}{2\pi}|t_{\mathbf{p,k}}|^2\Big[ \hat{g}^>_{L,\mathbf{k},11}(\omega-\Delta\mu)\hat{g}^<_{R,\mathbf{p},11}(\omega)-
 \hat{g}^<_{L,\mathbf{k},11}(\omega-\Delta\mu)\hat{g}^>_{R,\mathbf{p},11}(\omega)\Big].
\eeq
As pointed out in the previous section, the expression above contains $I_c$ and $I_1$.
First, $I_c$ arises from the contribution containing $\hat{g}^{<(>)}_{R,\mathbf{0},11}(\omega)=-\frac{2\pi\mu i}{g}\delta(\omega)$ and $t_{\mathbf{0,k}}$.
The corresponding tunneling element satisfies
\beq
t_{\mathbf{0,k}}=t_{c}k\delta_{\mathbf{k}_{\parallel},\mathbf{0}},
\eeq
where the Kronecker delta above represents the momentum conservation along the interface.
Thus, $I_{c}$ can analytically be calculated as
\beq
I_{c}=-\frac{4\pi^2\mu|t_{c}|^2\rho_{N}(-\Delta\mu) }{g},
\eeq
where $\rho_N$ is the local density of states in normal bosons.
On the other hand, for calculation of $I_1$, we utilize the following tunneling matrix expression:
\beq
t_{\mathbf{p,k}}=tpk\delta_{\mathbf{k}_{\parallel},\mathbf{p}_{\parallel}}.
\eeq
In a small bias regime in which the phonon approximation is reasonble~\cite{PhysRevA.64.033610,PhysRevResearch.2.023284}, $I_1$ obeys Ohm's law and is given by
\beq
I_1\approx\frac{T^{7/2}t^2\sqrt{2M^3}\mu\Delta\mu}{40\pi c^4}f_1(y),
\eeq
where
\beq
f_1(y)=\int_y^{\infty}dx\frac{x^3\sqrt{x-y}}{\sinh^2(x/2)},
\eeq
with $y=-\mu_0/T$. In Fig.~\ref{fig1},
we plot $f_1(y)$ which indicates that it reduced to a constant for $y\to0$ and experiences an exponential decay for $y\to\infty.$
\begin{figure}[htbp]
\centering
\includegraphics[width=5.5cm]{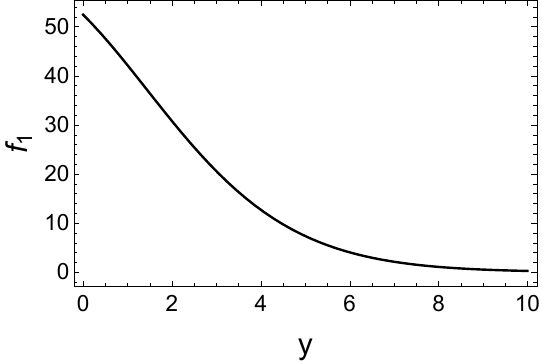}
\caption{\label{fig1} Behavior of $f_1(y)$.} 
\end{figure}

\subsection{Point contact case}
We next consider the point-contact system~\cite{PhysRevB.54.7366,PhysRevB.84.155414,husmann2015,PhysRevResearch.2.023284,ono2021} where 
the tunneling is assumed to occur at the origin in each reservoir, that is,
\beq
H_T=t[\phi^{\dagger}_L(\mathbf{0})\phi_R(\mathbf{0})+\phi^{\dagger}_R(\mathbf{0})\phi_L(\mathbf{0})].
\eeq 
The treatment of the above tunneling term is allowed for the single-mode quantum point contact whose width and length are so short that the 1D wire in detail is irrelevant. 
It is also straightforward to confirm that the above tunneling term is identical to one with momentum independent tunneling amplitude, i.e., $t_{\mathbf{p,k}} \to t.$

Correspondingly, the current can be expressed in terms of local Green's functions; $\hat{G}=\sum_{\mathbf{p,k}}\hat{G}_{\mathbf{p,k}}$ and $\hat{g}=\sum_{\mathbf{p}}\hat{g}_{\mathbf{p}}$.
Since the current expression contains full Green's functions, we now obtain them by using the Dyson equations.
By using the equation for $\hat{G}_{RR}$,
\beq
\hat{G}^{R(A)}_{RR}=\hat{g}^{R(A)}_R+\hat{g}_R\hat{t}\hat{g}^{R(A)}_L\hat{t}^{\dagger}\hat{G}^{R(A)}_{RR},
\eeq
we obtain
\beq
\hat{G}^{R(A)}_{RR}=\frac{1}{D^{R(A)}}\begin{pmatrix}
\hat{g}^{R(A)}_{R,11}-t^2\hat{g}^{R(A)}_{L,22}\{\hat{g}^{R(A)}_{R,11}\hat{g}^{R(A)}_{R,22}-(\hat{g}^{R(A)}_{R,12})^2 \}  & \hat{g}^{R(A)}_{R,12}\\
\hat{g}^{R(A)}_{R,12} & \hat{g}^{R(A)}_{R,22}-t^2\hat{g}^{R(A)}_{L,11}\{\hat{g}^{R(A)}_{R,11}\hat{g}^{R(A)}_{R,22}-(\hat{g}^{R(A)}_{R,12})^2 \}
\end{pmatrix},
\eeq
where
\beq
D^{R(A)}=1-t^2\{\hat{g}^{R(A)}_{R,11}\hat{g}^{R(A)}_{L,11}+\hat{g}^{R(A)}_{R,22}\hat{g}^{R(A)}_{L,22}\}+t^4\hat{g}^{R(A)}_{L,11}\hat{g}^{R(A)}_{L,22}\{\hat{g}^{R(A)}_{R,11}\hat{g}^{R(A)}_{R,22}-(\hat{g}^{R(A)}_{R,12})^2 \}.
\eeq
In addition, the Dyson equations for $\hat{G}_{LR(RL)}$ lead to
\beq
\hat{G}^{R(A)}_{RL}=\begin{pmatrix}
t\hat{G}^{R(A)}_{RR,11}\hat{g}^{R(A)}_{L,11} & t\hat{G}^{R(A)}_{RR,12}\hat{g}^{R(A)}_{L,22}\\
 t\hat{G}^{R(A)}_{RR,21}\hat{g}^{R(A)}_{L,11} & t\hat{G}^{R(A)}_{RR,22}\hat{g}^{R(A)}_{L,22} 
\end{pmatrix},
\eeq
\beq
\hat{G}^{R(A)}_{LR}=\begin{pmatrix}
t\hat{G}^{R(A)}_{RR,11}\hat{g}^{R(A)}_{L,11} & t\hat{G}^{R(A)}_{RR,12}\hat{g}^{R(A)}_{L,11}\\
 t\hat{G}^{R(A)}_{RR,21}\hat{g}^{R(A)}_{L,22} & t\hat{G}^{R(A)}_{RR,22}\hat{g}^{R(A)}_{L,22} 
\end{pmatrix}.
\eeq
By using above, it is straightforward to confirm 
\beq
[\hat{G}^A_{LR}]^{\dagger}=\hat{G}^R_{RL}.
\eeq

\begin{figure}[htbp]
\centering
\includegraphics[width=5.5cm]{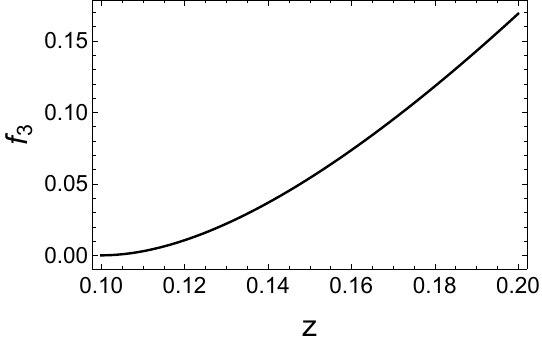}
\caption{\label{fig2} Behavior of $f_3(y,z)$, where $y=0.1$ is chosen.} 
\end{figure}
Based on the expressions of full Green's functions above, we next discuss the current expression in the point contact.
First, $I_{c}$ is calculated as
\beq
I_{c}=-\frac{2\pi\mu t^2\rho_{N}(-\Delta\mu)\Big[|1+t\hat{G}^R_{RL,11}(0)|^2+2t\text{Re}\{\hat{G}^A_{LR,21}(0)(1+t\hat{G}^R_{RL,11}(0))\}+t^2|\hat{G}^R_{RL,12}(0)|^2 \Big]}{g}.
\eeq
The remaining contributions are calculated as
\beq
I_1=4\pi^2 t^2\int_{-\infty}^{\infty} \frac{d\omega}{2\pi}|1+t\hat{G}^R_{RL,11}(\omega) |^2\hat{\rho}_{L,11}(\omega-\Delta\mu)\hat{\rho}_{R,11}(\omega)[n_L(\omega-\Delta\mu)-n_R(\omega)],
\eeq
\beq
I_2=8\pi^2 t^3\int_{-\infty}^{\infty} \frac{d\omega}{2\pi}\text{Re}\{\hat{G}^A_{LR,21}(\omega)(1+t\hat{G}^R_{RL,11}(\omega)) \}\hat{\rho}_{L,11}(\omega-\Delta\mu)\hat{\rho}_{R,12}(\omega)[n_L(\omega-\Delta\mu)-n_R(\omega)],
\eeq
\beq
I_3=4\pi^2 t^4\int_{-\infty}^{\infty} \frac{d\omega}{2\pi}|\hat{G}^R_{RL,12}(\omega) |^2\hat{\rho}_{L,11}(\omega-\Delta\mu)\hat{\rho}_{R,22}(\omega)[n_L(\omega-\Delta\mu)-n_R(\omega)],
\eeq
\beq
I_A=4\pi^2 t^4\int_{-\infty}^{\infty} \frac{d\omega}{2\pi}|\hat{G}^R_{RR,12}(\omega) |^2\hat{\rho}_{L,11}(\omega-\Delta\mu)\hat{\rho}_{L,22}(\omega+\Delta\mu)[n_L(\omega-\Delta\mu)-n_L(\omega+\Delta\mu)].
\eeq
As in the case of the planar junction, we evaluate each contribution at small $\Delta\mu$.
Except for $I_A$, the current contributions obey Ohm's law given by
\beq
I_1+I_2+I_3\approx G\Delta\mu,
\eeq
where
\beq
G=\frac{T^{3/2}M\sqrt{2M}\mu[t^2\{1-t^2\hat{g}^R_{L,11}(0)\hat{g}^R_{R,11}(0)\}^2-2t^4\hat{g}^R_{L,11}(0)\hat{g}^R_{R,12}(0)\{1-t^2\hat{g}^R_{L,11}(0)\hat{g}^R_{R,11}(0)\}+t^6(\hat{g}^R_{L,11}(0)\hat{g}^R_{R,12}(0))^2]f_2(y) }{16\pi^3c^3(D^R(0))^2},
\nonumber\\
\eeq
where
\beq
f_2(y)=\int_y^{\infty}dx\frac{x\sqrt{x-y}}{\sinh^2(x/2)}.
\eeq
As in the case of $f_1(y)$ appearing in the planar junction,  $f_2(y)$ converges to a constant for small  $y$ and is exponentially suppressed for large $y$.
On the other hand, $I_A$ is evaluated as
\beq
I_A\approx-\frac{T^2M^3t^4(\hat{g}^R_{R,12}(0))^2}{\pi^3(D^R(0))^2}f_3(y,z),
\eeq
where 
\beq
f_3(y,z)=\int_{-z+y}^{z-y}dx\sqrt{(z-y+x)(z-y-x)}\theta(z-y+x)\theta(z-y-x) \Big[\frac{1}{e^{x+z}-1}-\frac{1}{e^{x-z}-1}\Big],
\eeq
with $z=-\Delta\mu/T$.
Clearly, $f_3$ vanishes for $z<y$ and therefore $\Delta\mu<0$ as well as a small $y$ is necessary to have $I_A$ for low-energy transport.
The typical behavior of $f_3$ is shown in Fig.~\ref{fig2},  we have for the low-bias regime
\beq
f_3(y,z)\propto (z-y)^2\theta(z-y),
\eeq
that is, $I_A$ is proportional to $(\Delta\mu+\mu_0)^2$ and its proportional constant does not depend on temperature.

%\bibliography{reference}

\end{document}